\documentclass[12pt]{article}
\usepackage{amssymb}
\usepackage{amsfonts}
\usepackage{amsmath}
\usepackage{amsmath,amsthm}
\usepackage{graphicx,epsfig}

\def\be{\begin{equation}}
\def\ee{\end{equation}}
\begin{document}
\begin{center} {{\bf {Weak Gravitational Lensing of quantum perturbed Lukewarm Black Holes and cosmological constant effect}}\\
 \vskip 0.10 cm
 {{H. Ghaffarnejad\footnote{E-mail address: hghafarnejad@yahoo.com; hghafarnejad@profs.seman.ac.ir}} and M. A. Mojahedi \footnote{E-mail address:
 amirmojahed@semnan.ac.ir}}}
 \vskip 0.1 cm
 \textit{Faculty of Physics, Semnan
University, Semnan, Zip Code: 35131-19111, IRAN}
\end{center}
\begin{abstract}
Aim of the paper is study weak gravitational lensing of quantum
(perturbed) (QLBHL) and classical (CLBHL) Lukewarm black hole in
presence of cosmological parameter $\Lambda$. We apply numerical
method to evaluate deflection angle of bending light rays,  images
locations $\theta$ of sample source $\beta=-\frac{\pi}{4},$ and
corresponding magnifications $\mu.$
 There is not obtained real values for Einstein rings
locations $\theta_E(\beta=0)$ for CLBHL but we calculate them for
QLBHL. As experimental test of our calculations, we choose mass
$M$ of 60 type of most massive observed galactic black holes as
gravitational lens and study quantum matter fields effects on the
angle of bending light rays in presence of the cosmological
constant effects. We calculate locations of non-relativistic
images and corresponding magnifications. Numerical diagrams show
that the quantum matter effects cause to be reduce absolute values
of the quantum deflection angle with respect to the classical
ones. Sign of the quantum deflection angle is changed with respect
to the classical values in presence of the cosmological constant.
This means dominance of anti-gravity counterpart of the
cosmological horizon on the bending light rays angle with respect
to absorbing effects of local 60 type of observed most massive
black holes. Variations of the image positions and magnifications
are negligible by increasing dimensionless cosmological constant
$\epsilon=\frac{16\Lambda M^2}{3}.$ Deflection angle takes
positive (negative) values for CLBHL (QLBHL) and they decrease
very fast (slow) by increasing closest distance $x_0$ of bending
light ray and/or dimensionless cosmological parameter  for sample
giant black holes $0.001<\epsilon<0.01$.
\end{abstract}
\section{Introduction}
Gravitational deflection of light rays is one of the experimental
tests of general theory of relativity. This effect is observed in
presence of interstellar and large scale objects for instance
stars, black holes and clusters of galaxies [1, 2] ( see also [3])
and references therein). The gravitational lensing in weak field
limits
 can be used to measure masses of interstellar objects.  Micro-lensing is observed in cosmic sources
called as galactic micro-lensing. In the latter case there is made
only elementary and secondary images. Source and observer is
assumed to be located in asymptotically flat regions of the lens.
  Multiple relativistic images are formed via strong gravitational lensing where the
light ray moves very close to the black holes`s photon sphere and
circulates about the lenz (black hole) before than that arrives to
observer. In fact, exact analytic solutions of the null geodesics,
gravitational lens equations and deflection angle of light for the
Kerr-Newman and Kerr-Newman-(anti) de Sitter black holes have been
studied and derived for the first time in ref. [4]. The solutions
are expressed in terms of generalized hypergeometric functions of
Appell-Lauricella and Weierstrass elliptic functions. There all
the parameters of the theory mass, spin, electric charge and
cosmological constant enter the solutions on an equal footing (the
RNdS black hole is a special case of KNdS). In absence of
cosmological constant, influence of electric charge is studied by
Eiroa et al for RN black hole strong lensing [5]. They obtained
its relativistic image. Influence of the cosmological constant is
studied on the gravitational lensing by Sereno [6] in which
effects of the cosmological constant is negligible  near the lens
but not at distances far from the lens. In these large distances
the unresolved images are slightly de-magnified, radius of the
Einstein ring decreases and time delay increases. Gravitational
lensing of Kerr black hole is studied in weak field limits [7]
where the critical curves are still circles displaced from the
black hole location in the equatorial direction and the
corresponding caustic is point-like. In strong deflection limits
the Kerr black hole gravitational lensing is studied in ref. [8]
in which all observational quantities namely, image locations,
magnifications, and etc. are depended on the projection of the
spin on plan which is orthogonal to the line of sight.
Gravitational lensing is also studied from wormholes and naked
singularities. For instance strong deflection limits of
Janis-Newman-Winnicour (JNW) wormhole and Ellis wormhole
gravitational lensing are studied in ref. [9]. The JNW wormhole
exhibits the relativistic images, but not the Ellis wormhole due
to the absence of its photon sphere. Gravitational lensing from
rotating naked singularities is studied (see ref.[10]) in presence
of massless scalar fields in which scalar charge has important
effect on magnification signature (parity of the images) and
critical curves displacement but not on the point-like caustics.
In other word the point like caustics be moved away from the
optical axis without to affect the scalar charge. Locations of the
relativistic images and their separability for weakly naked
singularities are computed in the strong deflection limits by
using numerical method. Role of the scalar field is studied in
gravitational lensing
 [11] where a spherically symmetric static lens is characterized
by its mass and scalar charge parameters. Usually the nonlinear
electromagnetic fields of the black hole charge causes to
eliminate causal singularity of the black holes. In the latter
case they reach to regular supper-massive black holes located at
the center of galaxies. The first regular black hole was
introduced by Bardeen [12,13]. This is obtained from Einstein
field equation with nonlinear electromagnetic source [14,15]. In
this work we assume that a quantum perturbed Lukewram black hole
(QLBH) metric to be gravitational lens. Then we study its
gravitational lensing for weak deflection limits of bending light
rays. Metric of the QLBH was obtained [16] by solving the well
known backreaction metric equation where renormalized expectation
value of stress tensor operator of a massless quantum scalar field
is located in right side of the Einstein equation. It will be
quantum counterpart of source and causes to evaporate the RNdS
black hole which its initial ADM mass `M` is equal to its electric
charge. In ref. [16] we showed that its final state reaches to
remnant stable mini black hole.
 All observational quantities for instance locations of
non-relativistic images and magnifications are studied for QLBH
and compared with results of a CLBH lens. We will use numerical
method to solve the lens equation and calculation of corresponding
magnifications. As experimental test of our numerical calculations
we use 60 type of most massive observed galactic back holes as
gravitational lens and study their mass absorbing effect versus
the anti-gravity effects of cosmological constant on the quantum
(classical) deflection angle of bending light rays in the presence
(absence) of quantum scalar field effects.  The paper is organized as follows.\\
 In section 2 we call CLBH and QLBH metric from [16].
 In section 3 we use numerical method to evaluate deflection angle of bending light ray and plot its diagram against dimensionless
 cosmological parameter
 $\epsilon$.
In sections 4 we obtain elementary and secondary image locations .
In section 5 we evaluate magnifications of the non-relativistic
images  and plot their numerical diagrams. Section 6 denotes to
conclusion.
\section{Classical and quantum Lukewarm Black Hole}
Let us start with metric of final state of evaporating QLBH. This
metric was obtained previously in ref. [16] by solving the
backreaction equation via perturbation method as:
 \begin{equation}
(2M)^{-2}ds^2\cong-F_{B,C}(x)d\tau^2+\frac{dx^2}{F_{B,C}(x)}+x^2(d\theta^2+\sin^2\theta
d\varphi^2)
\end{equation}
\begin{equation} F_{B,C}(x)=1+\frac{1}{4
x^{2}}-\frac{\rho_{B,C}(x)}{x}-\epsilon\frac{ \sigma_{B,C}(x)
x^{2}}{4}
\end{equation}
where subscripts of $B,C$ denotes to the black hole and
cosmological horizons regions. We used units where  $c=G=\hbar=1$
and defined dimensionless quantities
\begin{equation}
\tau=\frac{t}{2M},~~~~x=\frac{r}{2M},~~~q=\frac{Q}{M}=1,~~~\epsilon=\frac{16\Lambda
M^2}{3}\end{equation} in which $Q$ denotes to the black hole
electric charge parameter. We will call $\epsilon$ to be
dimensionless cosmological parameter in what follows. Up to terms
in order of $\epsilon^2$ perturbation series expansion form of the
backreaction functions $\rho_{B,C}(x)$ and $\sigma_{B,C}(x)$ are
obtained by solving the well known backreaction equation
$G_{\mu\nu}=-8\pi \{T_{\mu\nu}^{EM}+<\hat{T}_{\mu\nu}>_{ren}\}$
where $T^{EM}_{\mu\nu}$ and $<\hat{T}_{\mu\nu}>_{ren}$ are
classical electromagnetic field stress tensor and renormalized
expectation value of quantum massless, charge-less scalar field
stress tensor operator respectively such that
\begin{equation} \rho_B(x)=\rho(x\to x_B)\simeq
\bigg\{\frac{U_1(x_B)}{(x-x_B)}-\frac{U_2(x_B)}{(x-x_B)^2}\bigg\}
\exp\bigg\{\frac{U_3(x_B)}{x-x_B}\bigg\},
\end{equation}
\begin{equation} \rho_C(x)=\rho(x\to
x_C)\simeq\bigg\{\frac{U_1(x_C)}{(x_C-x)}-\frac{U_2(x_C)}{(x_C-x)^2}\bigg\}
\exp\bigg\{\frac{U_3(x_C)}{x_C-x}\bigg\},
\end{equation}
\begin{equation}\sigma_B(x)=\sigma(x\to x_B)\simeq-\frac{4}{3\epsilon}
\bigg\{\frac{V_1(x_B)}{(x-x_B)}+\frac{V_2(x_B)}{(x-x_B)^2}\bigg\}
\exp\bigg\{\frac{V_3(x_B)}{x-x_B}\bigg\},\end{equation} and
\begin{equation} \sigma_C(x)=\sigma(x\to
x_C)\simeq-\frac{4}{3\epsilon}\bigg\{\frac{V_1(x_C)}{(x_C-x)}+\frac{V_2(x_C)}{(x_C-x)^2}\bigg\}
\exp\bigg\{\frac{V_3(x_C)}{x_C-x}\bigg\}\end{equation} where $x_B$
and $x_C$ are radius of the black hole and the cosmological event
horizon respectively given by
\begin{equation}
x_B=\frac{1}{2}+\frac{\sqrt{\epsilon
}}{8},~~~x_C=\frac{2}{\sqrt{\epsilon
}}-\frac{1}{2}-\frac{\sqrt{\epsilon }}{8}.
\end{equation}
Definition of the constants $U_{1,2,3}(x_{B,C})$ and
$V_{1,2,3}(x_{B,C})$ are given in the appendix A. CLBH metric is
obtained by solving $G_{\mu\nu}=-8\pi T^{EM}_{\mu\nu}$ (in absence
of backreaction corrections of Hawking radiation of quantum
massless scalar fields) such that $\rho_{B,C}(x)=1$ and
$\sigma_{B,C}(x)=1.$ Now, we use numerical method to calculate
deflection angle of bending light rays for both of QLBH and CLBH
metric as follows.
\section{Light deflection angle}
When the light ray moves at neighborhood of the black hole metric
(1) (the quantum lens), then deflects with angle [17]
\begin{equation}
\alpha=2\int_{x_{0}\geq x_{ps}}^{x_C(\epsilon)
}\frac{1}{x^{2}}\frac{dx}{\sqrt{1/
\tilde{b}^{2}-F_{B,C}(x,\varepsilon)/x^{2}}}-\pi
\end{equation}
where $x_{ps}$ is the photon sphere radius and $x_B$ and $x_C$ are
called radiuses of the black hole and the cosmological horizons
respectively given by (8). Here dimensionless impact parameter
$\tilde{b}=\frac{b}{2M}$ is given by
\begin{equation}\tilde{b}=\frac{b}{2M}=\frac{x_0}{\sqrt{F_{B,C}(x_0,\varepsilon)}}\end{equation}
in which impact parameter $b$ is defined in terms of constant
angular momentum $L$ and energy $E$ of the light ray as
$b=|\frac{L}{E}|.$ It is coordinate independent and so is
invariant of the system, same as the black hole mass $M$ and the
electric charge $Q$ . The photon sphere radius $x_{ps}$ is
obtained by largest positive root of the equation
\begin{equation} \frac{d}{dx}\left(\frac{x^2}{F_{B}(x)}\right)\big|_{x=x_{ps}}\approx0\end{equation}
where explicit form of the function
    $F_{B}(x,\epsilon)$ is obtained by inserting (4) and (6) into (2). The photon sphere equation (11) is
    obtained by solving the geodesic equation for a circular moving photon about the black hole center.
     Explicit form of the function $F_{C}(x,\epsilon)$ in the equation (10) is  obtained by inserting
    (7) and (9) into (2).
$x_0$ given in the integration (9) is closest approach distance
for which the bending light ray reaches to center of the lens. In
weak deflection limits of the gravitational lensing we must be
restrict the integration (9) to $x_0>x_{ps}$ for which
  $|\alpha|<\frac{3\pi}{2}\simeq4.7 (radian)$ and so we have not still relativistic images. Relativistic images are formed usually
   via circulation of bending light rays around the lens center
   for which deflection angle become $|\alpha|>\frac{3\pi}{2}$ (see figure 1 in ref. [18].
  For galactic large black hole we can approximate
$\epsilon\simeq10^{-22}$ experimentally (see [19]) for which the
equation (8) reads \be x_B\approx0.5,~~~~x_C\approx10^{11}.\ee
What  should we choose for particular value of $x_0$ if we use the
sample (12)? To do so we define boundary point $x_{b}$ via \be
F_B(x_b,\epsilon)=F_C(x_b,\epsilon)\ee  and \be
\frac{d}{dx}F_B(x_b,\epsilon)=\frac{d}{dx}F_C(x_b,\epsilon).\ee
The above equations do not give us analytic solution for $x_{b}$
against $\epsilon$ and so we must be obtain all possible values by
plotting their diagrams . Diagrams of the equations (11) and (13)
are plotted against $\epsilon$ in left panel of figure 1 where
solid line denotes to $x_{ps}(\epsilon)$ and dot line denotes to
$x_b(\epsilon)$. Right panel of the figure 1 describes diagram of
the equation (14). These 2 diagrams show that \be x_{ps}\approx
x_b<0.6,~~~~\epsilon<1\ee and so the functions (4) and (6) are
negligible when we calculate (9) by choosing the following
boundary conditions. \be 0.6\leq x_0\leq1.5,~~~~x_C\approx10^{11};
~~~\epsilon=10^{-22}\ee and/or \be
x_0=1.5,~~~~x_C=\frac{2}{\sqrt{\epsilon}}-\frac{1}{2}-\frac{\sqrt{\epsilon}}{8};~~~~~0<\epsilon<1.\ee
We should  point the dimensionless cosmological parameter given in
the Eq. (16) corresponds to the most massive black holes in the
central part of giant galaxies and certainly not for Milky Way.
Inserting (2), (5) and (7) as $F_C(x)=1+\frac{1}{4
x^{2}}-\frac{\rho_C(x)}{x}-\frac{\epsilon \sigma_C(x) x^{2}}{4}$
we plot diagram of  quantum and classical deflection angles
$\alpha_q$ and $\alpha_c$ respectively  defined by (9) against
$x_0$ ($\epsilon$) in the left (right) up panel of the figure 2
for  theoretical giant black holes sample $0.001<\epsilon<0.01$.
 Left-down panel in figure 2 denotes to numerical diagram of
the classical and quantum deflection angles of the bending light
rays for 60 type of experimental observed most massive galactic
black holes lens. We see that the weak  deflection limits
satisfies $|\alpha|<\frac{3\pi}{2}\simeq4.7$ in the figure 2.
Quantum deflection angle $\alpha_q$ is shown in this figure with
$Square$ symbol but classical deflection angle $\alpha_c$ is shown
with $+$ and/or $\times.$ Right-down panel in the figure 2 denotes
to schematic diagram of gravitational lensing setup. The classical
deflection angle is evaluated by inserting $F(x)=1+\frac{1}{4
x^{2}}-\frac{1}{x}-\frac{\epsilon x^{2}}{4},$ into the equation
(9) and integrating numerically for boundary conditions (16)
and/or (17)  corresponding to 60 type of observed most massive
black hole lens and theoretical regime $0.001<\epsilon<0.01$. We
see in the up left panel of the figure 2 that the deflection angle
decreases by increasing $x_0$ for CLBHL but not for QLBHL. In the
up left panel of the figure 2, we see that there are formed
relativistic images for CLBHL with $\alpha_c\to\infty$ and
$x_0\to1$ but not in case of QLBHL. We see in the up right panel
of the figure 2 that the deflection angle decreases faster
(slower) for CLBHL (QLBHL) by increasing the cosmological
parameter $\epsilon.$ Also these diagrams show that the quantum
deflection angle sign is changed with respect to the classical one
in the presence of the quantum matter field effects. Also absolute
value of quantum deflection angle become smaller than with respect
to the classical one in the presence of quantum matter fields
effects. In the next section we use the above results and seek
locations of non-relativistic images.
\section{Non-relativistic image locations}
There is proposed several  type of the gravitational lensing
equation  (see for instance [19,20]) where we choose
\begin{equation}
\tan\beta=\tan\theta-\frac{D_{ds}}{D_{s}}(\tan\theta+\tan(\alpha-\theta))
\end{equation}
where source angular location $\beta$ is made by crossing
observer-lens line and observer-source line (see down-right panel
in figure 2). Image angular location $\theta$ is made by crossing
observer-lens line and observer-image line. $D_{ds}$ denotes to
distance between lens and source. $D_{s}$ describes distance
between observer and source.  We set
$\frac{D_{ds}}{D_{s}}=\frac{1}{2}$ for simplification of the
problem. Applying  numerical method we solve (18), and plot its
diagram against $\epsilon$ for $\beta=0$ (Einstein rings)
 and sample source $\beta=-\frac{\pi}{4}$ for QLBHL. In case of CLBHL we do not obtain real values for Einstein rings location
 $\theta_E(\beta=0)$. Diagram of QLBHL Einstein rings is given in right panel of figure 3 where
 numerical values $\epsilon$  are chosen from mass of 60 type of observed most massive black holes which are in order
 of $\epsilon\approx10^{-22}$ satisfying the condition (16) and/or (17).
 We plot numerical diagrams of non-relativistic images locations for a sample source
 $\beta=-\frac{\pi}{4}$ from 60 type of experimental observed most massive galactic black holes lens
  in left (right) down panel of figure 4 for CLBHL (QLBHL).
  For theoretical sample  CLBHL (QLBHL) lens $0.001<\epsilon<0.01,$ diagrams of images locations of the source $\beta=-\frac{\pi}{4}$ is given
  in left (right) up panel of  figure 4. All diagrams show that the anti-gravity effects of the dimensionless cosmological
 constant $\epsilon$ is dominant with respect to local absorbing property of the supper massive black holes because all $\theta(\epsilon)$
 diagrams treat
 as constant function versus $\epsilon.$
\section{Magnification}
Applying Liouville's theorem, the gravitational lensing causes to
preserves surface brightness but it does change the apparent solid
angle of a source. This change is evaluated via magnification
$\mu$. The amount of magnification is given by ratio of the image
area to the source area and so it is a dimensionless numerical
quantity. Usually it is obtained larger than one in size but when
its amount become less than one then it refers to a reduction in
size which sometimes called as "de-magnification". Magnification
formula is given by
\begin{equation}
\mu=\mu_t\mu_r=\left(\frac{\sin\theta}{\sin\beta}\right)\left(\frac{d\theta}{d\beta}\right)
\end{equation}
 where $\mu_t$ and $\mu_r$ defined by   \begin{equation}\mu_{t}=\frac{\sin\theta}{\sin\beta},~~~~\mu_{r}=\frac{d\theta}{d\beta}
\end{equation}
are called  tangential and radial magnifications respectively.
Eliminating $\beta$ via (18) and assuming
$\frac{D_{d,s}}{D_s}=\frac{1}{2}$ the equations (19) and (20)
become respectively
\begin{equation}
\mu^{\pm}(\theta)=\frac{\pm2 \csc \left(\tan
^{-1}\left(\frac{1}{2} (\tan
(\theta-\alpha)+\tan\theta)\right)\right)
   }{\sec ^2(\alpha -\theta )+\sec ^2\theta}
   $$$$
   \times\sec ^2\left(\tan ^{-1}\left(\frac{1}{2} (\tan (\theta-\alpha )+\tan\theta)\right)\right) \sin
   \theta
\end{equation}
and
 \begin{equation}
\mu_{r}(\theta)=\frac{2 \sec ^2\left(\tan ^{-1}\left(\frac{1}{2}
(\tan (\theta-\alpha)+\tan \theta
   )\right)\right)}{\sec ^2(\alpha -\theta )+\sec ^2\theta},
 \end{equation}
 \begin{equation}
\mu^{\pm}_{t}(\theta)=\pm\csc \left[\tan ^{-1}\left(\frac{1}{2}
(\tan (\theta-\alpha)+\tan\theta)\right)\right]\sin
   \theta.
 \end{equation}
 In case of CLBHL we do not obtain real values for Einstein rings
location
 $\theta_E(\beta=0)$ and so their magnifications will not have real values. Diagrams of QLBHL Einstein
  rings magnifications are given in left panel of figure 3 where
 numerical values $\epsilon$ are chosen from mass of 60 type of observed most massive black holes which are in order of $\epsilon\approx10^{-22}$.
 In figure 5, we plot numerical diagrams of non-relativistic images magnifications of a sample source
 $\beta=-\frac{\pi}{4}$ from 60 type of experimental observed most massive galactic black holes with $\epsilon\approx10^{-22}$ and
 for theoretical sample $0.001<\epsilon<0.01.$ Diagrams show raising of magnifications by increasing dimensionless cosmological constant
 $\epsilon$ for regime  $0.001<\epsilon<0.01$ (see up-right panel in figure 5) but show (approximately) constant magnification  for observable regime
 $\epsilon\approx10^{-12}$ (see down-right panel in figure 5).\\
 As a future work we would like to consider gravitational lensing
of quantum perturbed rotating black hole background which is a
more realistic background. Casals et al [21] are studied quantum
fields on a rotating BTZ (Banados, Teitelboim and Zanelli) black
hole and obtained quantum perturbed metric in analytic form
recently. For given values of black hole mass and angular momentum
they obtained that the quantum effects lead to a growth of both
the event horizon and the radius of the ergo-sphere, reducing the
angular momentum compared to the unperturbed values. Also quantum
effects give rise to the formation of a curvature singularity at
the Cauchy horizon but show no evidence of a super-radiant
instability. BTZ black hole metric solution is obtained in 1992
[22], where it came as a surprise because it is believed that no
black hole solutions are shown to exist for a negative
cosmological constant. BTZ black hole which is a three dimensional
black hole has remarkably similar properties to the 3+1
dimensional black hole, which would exist in our real universe.
When the cosmological constant is zero, a vacuum solution of
(2+1)-dimensional gravity is necessarily flat (the Weyl tensor
vanishes in three dimensions, while the Ricci tensor vanishes due
to the Einstein field equations, so the full Riemann tensor
vanishes), and it can be shown that no black hole solutions with
event horizons exist. By introducing dilatons, we can have black
holes. We do have conical angle deficit solutions, but they don't
have event horizons. It therefore came as a surprise when black
hole solutions were shown to exist for a negative cosmological
constant. The BTZ 3-dimensional black hole has properties similar
to the ordinary black holes in 3+1 dimensions as follows: (a) It
satisfies the `no hair theorem` (b) It has the same
thermodynamical properties, namely its entropy is captured by a
law directly analogous to the Bekenstein bound in
(3+1)-dimensions, essentially with the surface area replaced by
the BTZ black hole's circumference. (c) Like the Kerr black hole,
a rotating BTZ black hole contains an inner and an outer horizon
together with ergosphere. BTZ black hole could arise from
collapsing matter and its energy-moment tensor is calculated as
same as (3+1) black holes In ref. [23]. In short , the BTZ black
holes without any electric charge are locally
isometric to anti-de Sitter space.\\
However we should use our quantum framework to obtain physical
effects of the Kerr and/or the BTZ rotating black holes angular
momentum on the gravitational lensing of light rays which can be
choose as next challenge in this subject. At least we can tell
some physical interpretation about the black hole lens angular
momentum on the gravitational lensing as follows: All rotating
black hole have ergosphere region which treats as particle
accelerators. Ejecting particles from ergosphere to regions far
from the black hole photon sphere region affect on moving photons
and so deviate their velocity and direction which in the post
Newtonian limits they do because of rotating black holes Coriolis
acceleration. These affect finally on images locations and their
magnifications.\\ Now we call 60 type of supper massive observed
galactic black holes mass to obtain quantum characteristics for
the gravitational lensing given in the tables 1-14 as follows.\\
\textbf{List of observed supper massive black holes,
$M_{\bigodot}\cong1.99\times10^{30} kg$}
\begin{enumerate}
    \item S5 0014+81, $M/M_{\bigodot}=4\times10^{10}$ ([24,25,26])
    \item SDSS J102325.31+514251.0,
    $M/M_{\bigodot}=3.31\times10^{10}$ ([27])
    \item Black hole of central quasar of H1821+643,
    $M/M_{\bigodot}=3\times10^{10}$ ([28])
    \item APM 08279+5255, $M/M_{\bigodot}=2.3\times10^{10}$ ([29])
    \item NGC 4889, $M/M_{\bigodot}=2.1\times10^{10}$ ([30])
    \item Black hole of central elliptical galaxy of Phoenix
    Cluster, $M/M_{\bigodot}=2.0\times10^{10}$ ([31])
    \item SDSS J074521.78+734336,
    $M/M_{\bigodot}=1.95\times10^{10}$ ([27])
    \item OJ 287 primary, $M/M_{\bigodot}=1.8\times10^{10}$ ([32])
    \item NGC 1600, $M/M_{\bigodot}=1.7\times10^{10}$ ([33])
    \item SDSS J08019.69+373047.3,
    $M/M_{\bigodot}=1.514\times10^{10}$ ([27])
    \item SDSS J115954.33+201921.1,
    $M/M_{\bigodot}=1.412\times10^{10}$ ([27])
    \item SDSS J075303.34+423130.8,
    $M/M_{\bigodot}=1.38\times10^{10}$ ([27])
    \item SDSS J080430.56+542041.1,
    $M/M_{\bigodot}=1.35\times10^{10}$ ([27])
    \item SDSS J081855.77+095848.0,
    $M/M_{\bigodot}=1.2\times10^{10}$ ([27])
    \item SDSS J0100+2802, $M/M_{\bigodot}=1.2\times10^{10}$ ([34,54])
    \item SDSS J082535.19+512706.3,
    $M/M_{\bigodot}=1.122\times10^{10}$ ([27])
    \item SDSS J013127.34-032100.1,
    $M/M_{\bigodot}=1.1\times10^{10}$ ([35])
    \item Black hole of central elliptical galaxy of MS
    0735.6+7421, $M/M_{\bigodot}=1.0\times10^{10}$ ([36])
    \item PSO J334.2028+01.4075,
    $M/M_{\bigodot}=1.0\times10^{10}$ ([37])
    \item Black hole of central elliptical galaxy of RX
    J1532.9+3021, $M/M_{\bigodot}=1.0\times10^{10}$ ([38])
    \item QSO B2126-158, $M/M_{\bigodot}=1.0\times10^{10}$ ([26])
    \item Holmberg 15A, $M/M_{\bigodot}=1.0\times10^{10}$ ([39])
    \item SDSS J015741.57-010629.6,
    $M/M_{\bigodot}=9.8\times10^{9}$ ([27])
    \item NGC 3842,Brightest galaxy in the Leo Cluster,
    $M/M_{\bigodot}=9.7\times10^{9}$ ([30])
    \item SDSS J230301.45-093930.7,
    $M/M_{\bigodot}=9.12\times10^{9}$ ([27])
    \item SDSS J075819.70+202300.9,
    $M/M_{\bigodot}=7.8\times10^{9}$ ([27])
    \item CID-947, $M/M_{\bigodot}=7.0\times10^{9}$ ([40])
    \item SDSS J080956.02+502000.9,
    $M/M_{\bigodot}=6.45\times10^{9}$ ([27])
    \item SDSS J014214.75+002324.2,
    $M/M_{\bigodot}=6.31\times10^{9}$ ([27])
    \item Messier 87, $M/M_{\bigodot}=6.30\times10^{9}$ ([41])
    \item SDSS J025905.63+001121.9,
    $M/M_{\bigodot}=5.25\times10^{9}$ ([27])
    \item SDSS J094202.04+042244.5,
    $M/M_{\bigodot}=5.13\times10^{9}$ ([27])
    \item QSO B0746+254, $M/M_{\bigodot}=5.0\times10^{9}$ ([26])
    \item QSO B2149-306, $M/M_{\bigodot}=5.0\times10^{9}$ ([26])
    \item NGC 1277, $M/M_{\bigodot}=5.0\times10^{9}$ ([42])
    \item SDSS J090033.50+421547.0,
    $M/M_{\bigodot}=4.7\times10^{9}$ ([27])
    \item Messier 60, $M/M_{\bigodot}=4.5\times10^{9}$ ([43])
    \item SDSS J011521.20+152453.3,
    $M/M_{\bigodot}=4.1\times10^{9}$ ([27])
    \item QSO B0222+185, $M/M_{\bigodot}=4.0\times10^{9}$ ([26])
    \item Hercules A (3C 348),
    $M/M_{\bigodot}=4.0\times10^{9}$ ([44])
    \item SDSS J213023.61+122252.0,
    $M/M_{\bigodot}=3.5\times10^{9}$ ([27])
    \item SDSS J173352.23+540030.4,
    $M/M_{\bigodot}=3.4\times10^{9}$ ([27])
    \item SDSS J025021.76-075749.9,
    $M/M_{\bigodot}=3.1\times10^{9}$ ([27])
    \item SDSS J030341.04-002321.9,
    $M/M_{\bigodot}=3.0\times10^{9}$ ([27])
    \item QSO B0836+710,  $M/M_{\bigodot}=3.0\times10^{9}$ ([26])
    \item SDSS J224956.08+000218.0,
    $M/M_{\bigodot}=2.63\times10^{9}$ ([27])
    \item SDSS J030449.85-000813.4,
    $M/M_{\bigodot}=2.4\times10^{9}$ ([27])
    \item SDSS J234625.66-001600.4,
    $M/M_{\bigodot}=2.24\times10^{9}$ ([27])
    \item ULAS J1120+0641, $M/M_{\bigodot}=2.0\times10^{9}$ ([45,46])
    \item QSO 0537-286, $M/M_{\bigodot}=2.0\times10^{9}$ ([26])
    \item NGC 3115, $M/M_{\bigodot}=2.0\times10^{9}$ ([47])
    \item Q0906+6930, $M/M_{\bigodot}=2.0\times10^{9}$ ([48])
    \item QSO B0805+614, $M/M_{\bigodot}=1.5\times10^{9}$ ([26])
    \item Messier 84, $M/M_{\bigodot}=1.5\times10^{9}$ ([49])
    \item QSO B225155+2217, $M/M_{\bigodot}=1.0\times10^{9}$([26])
    \item QSO B1210+330, $M/M_{\bigodot}=1.0\times10^{9}$([26])
    \item NGC 6166, Central galaxy of Abell 2199,
    $M/M_{\bigodot}=1.0\times10^{9}$ ([50])
    \item Cygnus A, $M/M_{\bigodot}=1.0\times10^{9}$ ([51])
    \item Sombrero Galaxy, $M/M_{\bigodot}=1.0\times10^{9}$ ([52])
    \item Messier 49, $M/M_{\bigodot}=1.0\times10^{9}$([53])
\end{enumerate}
\section{Conclusions}
In this paper we studied weak gravitational lensing of QLBL and
CLBL in presence of the dimensionless cosmological constant
parameter $\epsilon$. In weak deflection limits we obtained
diagrams of deflection angle integral,
  non-relativistic image locations and corresponding magnifications against $\epsilon$ as numerically.
As experimental result of our work we choose 60 type of most
massive observed black hole mass to be experimental gravitational
lens and obtain deflection angle of bending light rays, their
non-relativistic image locations and magnifications in both of
CLBHL and QLBHL regime. Our work predicts dominance of
anti-gravity property of the cosmological constant effect versus
the mass absorbing effects of most massive galactic black holes on
the weak gravitational lensing of the bending light rays. Sign of
the quantum deflection angle is changed with respect to the
classical deflection angle. Absolute value of quantum deflection
angle reduces with respect to the classical deflection angle of
bending light rays for all values of dimensionless cosmological
constant. As a future work we will proceed to study our quantum
frame work on rotating most massive galactic black holes as same
as Kerr and/or BTZ which are more realistic background. Ejecting
particles from their ergosphere interfere moving photons around
the black hole photon sphere and so deviate their velocity and
direction.
   \begin{figure}
   \centering
   \includegraphics[width=6cm]{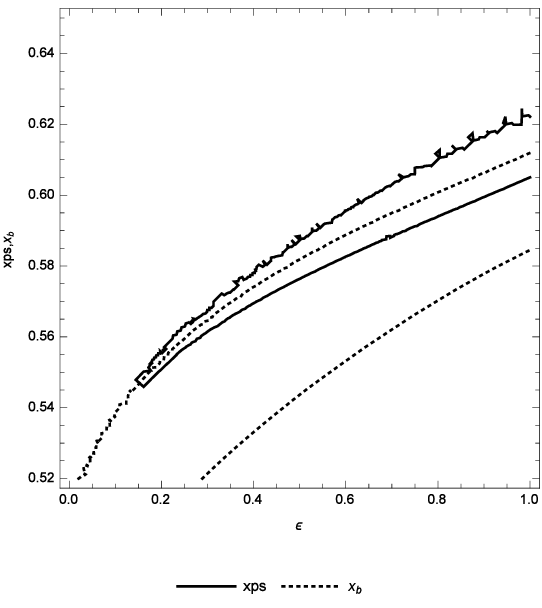}
   \includegraphics[width=6cm]{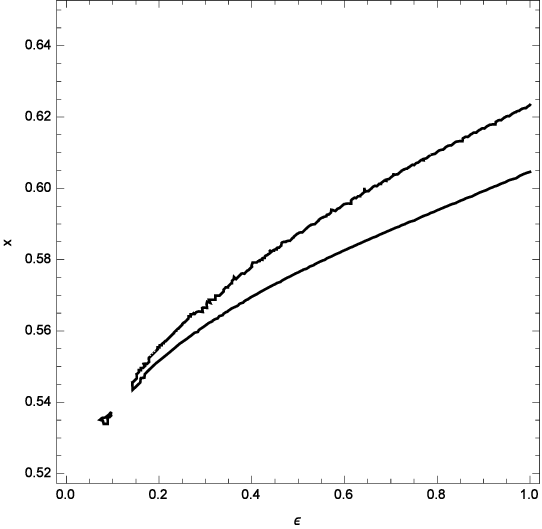}
   \caption{Left panel shows diagram of the equations (11) and (13) plotted against $\epsilon$ with solid and dot lines
 respectively. Right panel shows diagram of the equation (14)
 plotted against $\epsilon$. }
   \label{Fig1}
   \end{figure}
   \begin{figure}
   \centering
    \includegraphics[width=6cm]{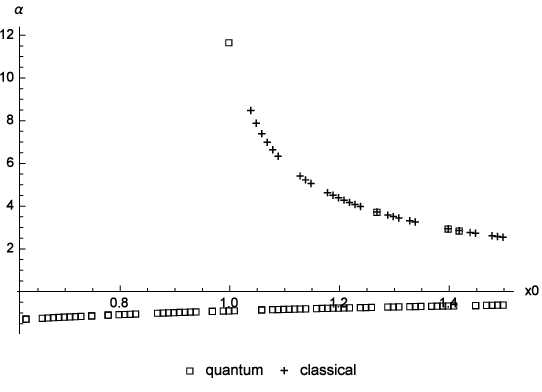}
   \includegraphics[width=6cm]{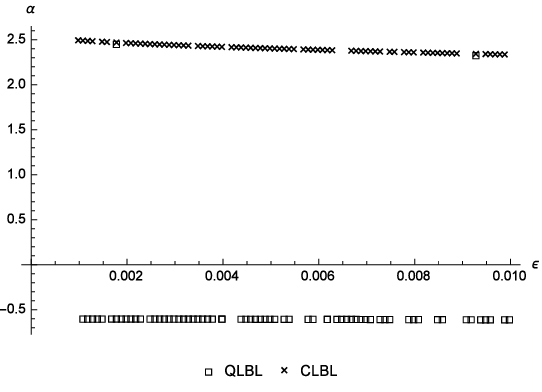}
    \includegraphics[width=9cm]{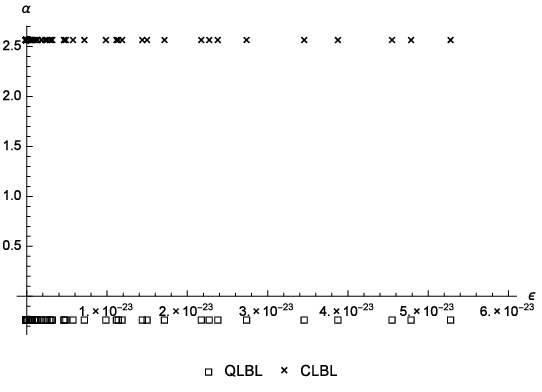}
     \includegraphics[width=4cm]{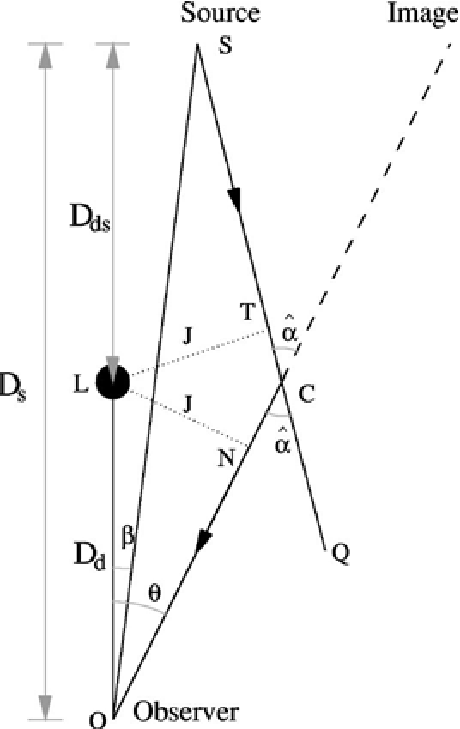}
  \caption{Diagram of the deflection angle (9) is plotted against $x_0$ for
$\varepsilon=10^{-22}, 0.6<x_0<1.5$ (up-left panel) and against
$\epsilon$ for $x_0=1.5;0.001<\epsilon<0.01$ (up-right panel).
Diagram of the bending angle $\alpha$ is plotted for the CLBHL and
QLBHL with + and $Square$ symbols respectively. Down left-panel
denotes to diagram of $\alpha$ plotted against $\epsilon$ for 60
type  of experimental observed large black holes. Down-right-panel
denotes to gravitational lensing setup.}
   \label{Fig2}
   \end{figure}
\begin{figure}
  \vspace{2mm}
   \begin{center}
  \includegraphics[width=6cm]{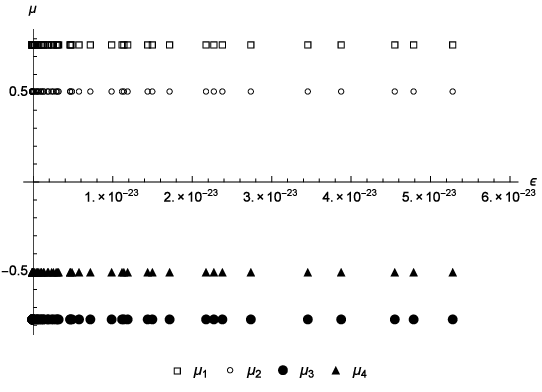}
  \includegraphics[width=6cm]{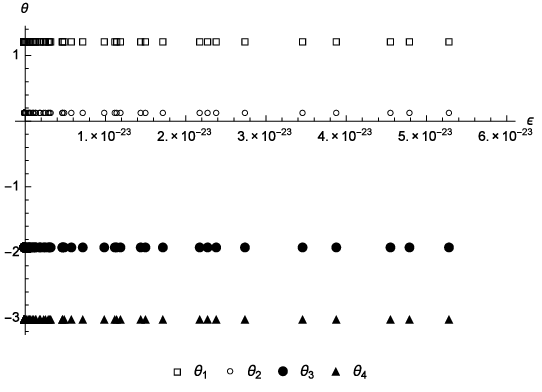}
   \caption{ Angular
   locations $\theta_{1,2,3,4}$ and corresponding magnifications $\mu_{1,2,3,4}$ of Einstein rings $(\beta=0)$
   for QLBHL and we have not obtain real values for CLBHL. }
   \label{Fig:lightcurve-ADAri}
   \end{center}
\end{figure}
\begin{figure}
  \begin{center}
\includegraphics[width=6cm]{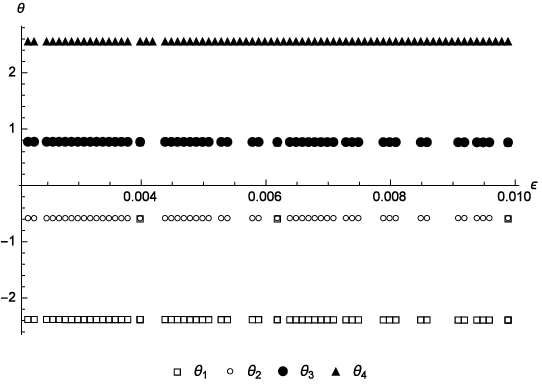}
  \includegraphics[width=6cm]{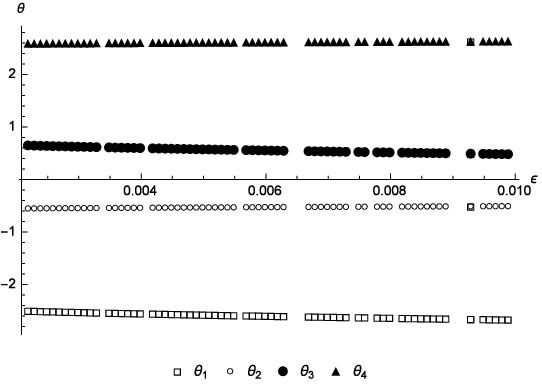}
 \includegraphics[width=6cm]{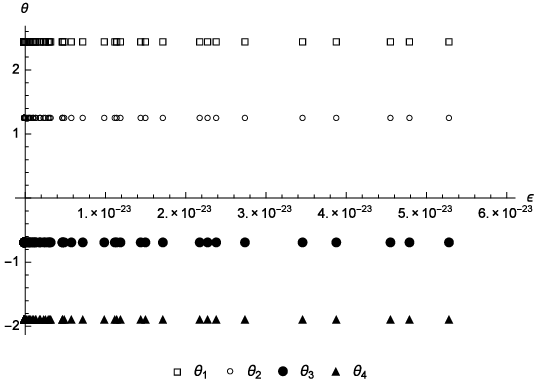}
  \includegraphics[width=6cm]{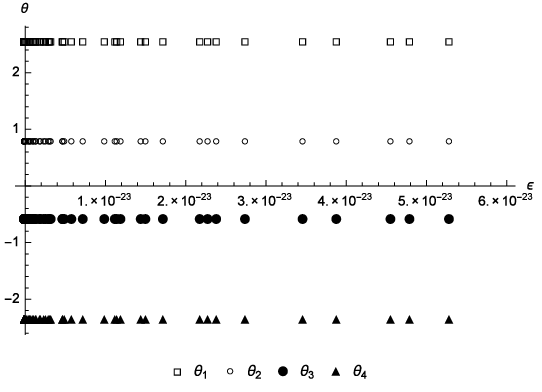}
  \caption{Upper panels are image angular locations $\theta_{1,2,3,4}$ of sample source $\beta=-\frac{\pi}{4}$  plotted against $\epsilon$
 for QLBHL (left panel) and CLBHL (right panel) respectively. Lower panels are image angular locations for 60 kind observed large
  black holes as QLBHL (left) and CLBHL (right).}
  \end{center}
\end{figure}
\begin{figure}
\includegraphics[width=6cm]{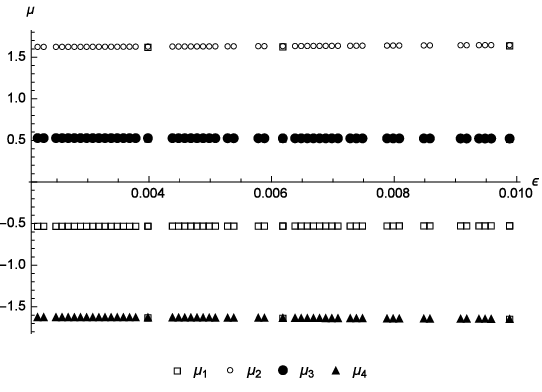}
 \includegraphics[width=6cm]{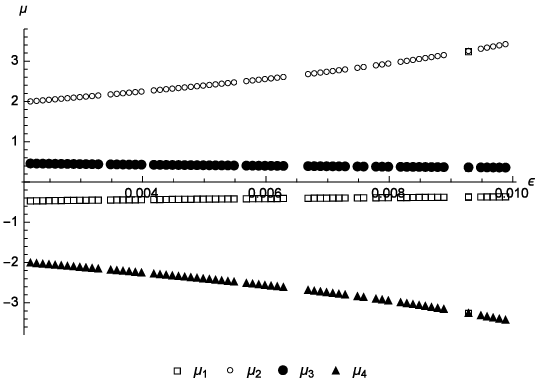}
 \includegraphics[width=6cm]{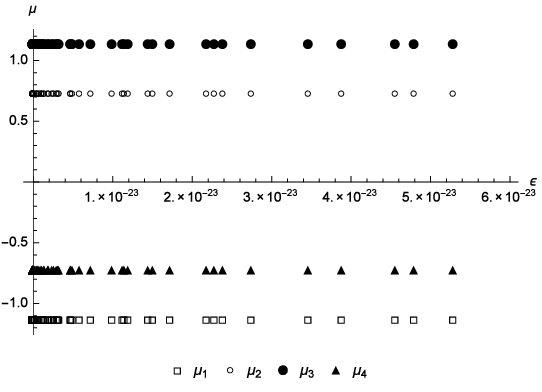}
 \includegraphics[width=6cm]{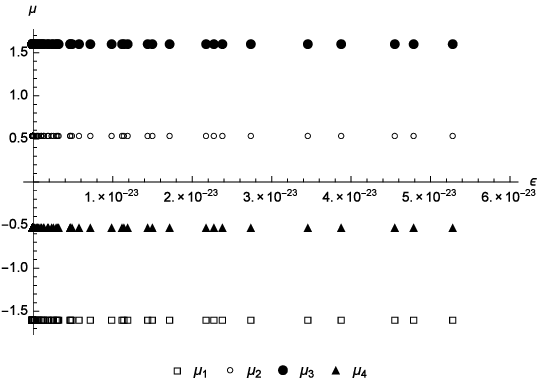}
 \caption{Upper panel is diagrams of
magnification of images positions $\theta_{1,2,3,4}$ of sample
source $\beta=-\frac{\pi}{4}$ is plotted against $\epsilon$ for
QLBHL (left panel) and CLBHL (right panel) respectively. Down left
(right)-panel denotes to diagram of magnification $\mu_{1,2,3,4}$
of non-relativistic images plotted against $\epsilon$ for 60 kind
of experimental observed large black holes as QLBHL (CLBHL).}
 \label{Fig:plot2}
\end{figure}
\section{Appendix}
 The constants $U_{1,2,3}(x_{B,C})$ and $V_{1,2,3}(x_{B,C})$ are calculated in ref.
 [16]
  as follows. \begin{equation} U_1(x_B)=\frac{\pi}{12}+\frac{4x_B^4[A(x_B)+B(x_B)/6]}{3}-\frac{16\pi
M^2C_2x_B^4}{3}
 \end{equation}
\begin{equation} U_2(x_B)=\frac{2\pi x_B^5[A(x_B)-8\pi
M^2C_2]}{3G(x_B)}
 \end{equation}
 \begin{equation} U_3(x_B)=2x_B^3[A(x_B)-8\pi
M^2C_2]
 \end{equation}
\begin{equation} U_1(x_C)=\pi/24+2x_C^4[A(x_C)+B(x_C)/6]/3-8\pi
M^2C_2x_C^4/3
 \end{equation}
\begin{equation} U_2(x_C)=\frac{\pi
x_C^3(1+4x_C^2)}{2G(x_C)}[\frac{\pi}{4}+4A(x_C)+\frac{2B(x_C)}{3}-16\pi
M^2C_2][A(x_C)-8\pi M^2C_2]
 \end{equation}
 \begin{equation} U_3(x_C)=-\frac{x_C^3[A(x_C)-8\pi
M^2C_2]}{2}
 \end{equation}
 \begin{equation} V_1(x_B)=\pi/3x_B^3+4x_B[A(x_B)+B(x_B)/6]-16\pi
M^2 x_BC_2
 \end{equation}
\begin{equation} V_2(x_B)=\frac{2\pi x_B^2[A(x_B)-8\pi
M^2C_2]}{G(x_B)}
 \end{equation}
 \begin{equation} V_3(x_B)=2x_B^3[A(x_B)-8\pi
M^2C_2]
 \end{equation}
\begin{equation} V_1(x_C)=\pi/8x_C^3+2x_C[A(x_C)+B(x_C)/6]-8\pi
M^2C_2x_C
 \end{equation}
\begin{equation} V_2(x_C)=\frac{\pi
x_C^2(1+4x_C^2)[A(x_C)-8\pi
M^2C_2]}{2G(x_C)}[\frac{\pi}{4x_C^2}+4A(x_C)+\frac{2B(x_C)}{3}-16\pi
M^2C_2]
 \end{equation}
 \begin{equation} V_3(x_C)=-\frac{x_C^3[A(x_C)-8\pi
M^2C_2]}{2}
 \end{equation}
where  $A(x,\varepsilon)$,$B(x,\varepsilon)$,$G(x,\varepsilon)$
and $C_{2}$ is defined as
\begin{equation}
A[x,\varepsilon]=\frac{1}{24 x^6}-\frac{1}{4 x^5}+\frac{1}{2
x^4}-\frac{1}{3 x^3}-\frac{\epsilon }{8
   x^2}+\frac{\epsilon }{4 x}
\end{equation}
\begin{equation}
B[x,\varepsilon]=\frac{3}{8 x^6}-\frac{2}{x^5}+\frac{7}{2
x^4}-\frac{2}{x^3}+\frac{x^2 \epsilon ^2}{8}-\frac{\epsilon
   }{2 x^2}+\frac{\epsilon }{x}-\frac{\epsilon }{2}
   \end{equation}
\begin{equation}
G[x,\varepsilon]=\pi-64 \pi  C_2 M^{2} x^{2}+16 x^{2}[A(x,\epsilon
)+\frac{1}{6} B(x,\epsilon )]
\end{equation}
\begin{equation}
C_{2}= \frac{18 \left(3 A\left(x_B,\epsilon
\right)-A\left(x_C,\epsilon \right)\right)+5
   B\left(x_B,\epsilon \right)-B\left(x_C,\epsilon \right)}{240 \pi  M^{2}}
\end{equation}
\center Table 1: Quantum Lensing characteristics for observed most
massive black holes 1-5, $\mu_4=-\mu_2,\mu_3=-\mu_1$
\begin{tabular}{|c|c|c|c|c|c|}
\hline
 $Black~Hole$ & 1 & 2 & 3& 4 & 5 \\
  \hline
 $\epsilon $ & $1.9\times 10^{-22}$ & $1.3\times 10^{-22}$ & $1.1\times 10^{-22}$ & $6.3\times
   10^{-23}$ & $5.3\times 10^{-23}$ \\
 $x_b$ & $0.5$ & $0.5$ & $0.5$ & $0.5$ & $0.5$ \\
 $x_c$ & $1.4\times 10^{11}$ & $1.7\times 10^{11}$ & $1.9\times 10^{11}$ & $2.5\times 10^{11}$ &
   $2.7\times 10^{11}$ \\
 $A(x_b)$ & $-8.9\times 10^{-16}$ & $-4.4\times 10^{-16}$ & $-8.9\times 10^{-16}$ &
   $-4.4\times 10^{-16}$ & $-4.4\times 10^{-16}$ \\
 $A(x_c)$ & $2.2\times 10^{-34}$ & $1.3\times 10^{-34}$ & $9.4\times 10^{-35}$ &
   $4.2\times 10^{-35}$ & $3.2\times 10^{-35}$ \\
 $B(x_b)$ & $-9.6\times 10^{-23}$ & $-6.6\times 10^{-23}$ & $-5.4\times 10^{-23}$ &
   $-3.2\times 10^{-23}$ & $-2.6\times 10^{-23}$ \\
 $B(x_c)$ & $-2.5\times 10^{-38}$ & $-1.\times 10^{-38}$ & $-1.3\times 10^{-38}$ &
   $-7.\times 10^{-39}$ & $3.4\times 10^{-39}$ \\
\hline
\end{tabular}
\center Table 2: Quantum Lensing characteristics for observed most
massive black holes 6-10, $\mu_4=-\mu_2,\mu_3=-\mu_1$
\begin{tabular}{|c|c|c|c|c|c|}
\hline $Black~Hole$ & 6 & 7 & 8 & 9 & 10 \\
  \hline
 $\epsilon $ & $4.8\times 10^{-23}$ & $4.6\times 10^{-23}$ & $3.9\times 10^{-23}$ & $3.5\times
   10^{-23}$ & $2.8\times 10^{-23}$ \\
 $x_b$ & $0.5$ & $0.5$ & $0.5$ & $0.5$ & $0.5$ \\
 $x_c$ & $2.9\times 10^{11}$ & $3.\times 10^{11}$ & $3.2\times 10^{11}$ & $3.4\times 10^{11}$ &
   $3.8\times 10^{11}$ \\
 $A(x_b)$ & $-4.4\times 10^{-16}$ & $-8.9\times 10^{-16}$ & $-4.4\times 10^{-16}$ &
   $-4.4\times 10^{-16}$ & $-8.9\times 10^{-16}$ \\
 $A(x_c)$ & $2.8\times 10^{-35}$ & $2.6\times 10^{-35}$ & $2.\times 10^{-35}$ &
   $1.7\times 10^{-35}$ & $1.2\times 10^{-35}$ \\
 $A(x_b)$ & $-2.4\times 10^{-23}$ & $-2.3\times 10^{-23}$ & $-1.9\times 10^{-23}$ &
   $-1.7\times 10^{-23}$ & $-1.4\times 10^{-23}$ \\
 $A(x_c)$ & $-7.5\times 10^{-39}$ & $-1.\times 10^{-39}$ & $5.7\times 10^{-40}$ &
   $2.\times 10^{-40}$ & $-1.1\times 10^{-39}$ \\
\hline
\end{tabular}
\center Table 3: Quantum Lensing characteristics for observed most
massive black holes 11-15, $\mu_4=-\mu_2,\mu_3=-\mu_1$
\begin{tabular}{|c|c|c|c|c|c|}
 \hline $Black~Hole$ & 11 & 12 & 13 & 14 & 15 \\
  \hline
 $\epsilon $ & $2.4\times 10^{-23}$ & $2.3\times
   10^{-23}$ & $2.2\times 10^{-23}$ &
   $1.7\times 10^{-23}$ & $1.5\times
   10^{-23}$ \\
 $x_b$ & $0.5$ & $0.5$ & $0.5$ & $0.5$ & $0.5$
   \\
 $x_c$ & $4.1\times 10^{11}$ & $4.2\times
   10^{11}$ & $4.3\times 10^{11}$ &
   $4.8\times 10^{11}$ & $5.1\times
   10^{11}$ \\
 $A(x_b)$ & $-4.4\times 10^{-16}$ &
   $-4.4\times 10^{-16}$ & $-8.9\times
   10^{-16}$ & $-4.4\times 10^{-16}$ &
   $-4.4\times 10^{-16}$ \\
 $A(x_c)$ & $9.8\times 10^{-36}$ &
   $9.1\times 10^{-36}$ & $8.5\times
   10^{-36}$ & $6.\times 10^{-36}$ &
   $4.9\times 10^{-36}$ \\
 $A(x_b)$ & $-1.2\times 10^{-23}$ &
   $-1.1\times 10^{-23}$ & $-1.1\times
   10^{-23}$ & $-8.6\times 10^{-24}$ &
   $-7.6\times 10^{-24}$ \\
 $A(x_c)$ & $-9.5\times 10^{-40}$ &
   $2.4\times 10^{-39}$ & $-5.\times
   10^{-40}$ & $7.1\times 10^{-40}$ &
   $-1.3\times 10^{-39}$ \\
\hline
\end{tabular}
\center Table 4: Quantum Lensing characteristics for observed most
massive black holes 16-20, $\mu_4=-\mu_2,\mu_3=-\mu_1$
\begin{tabular}{|c|c|c|c|c|c|}
\hline $Black~Hole$ & 16 & 17 & 18 & 19 & 20 \\
\hline
 $\epsilon $ & $1.5\times 10^{-23}$ & $1.2\times
   10^{-23}$ & $1.2\times 10^{-23}$ &
   $1.1\times 10^{-23}$ & $1.\times
   10^{-23}$ \\
 $x_b$ & $0.5$ & $0.5$ & $0.5$ & $0.5$ & $0.5$
   \\
 $x_c$ & $5.2\times 10^{11}$ & $5.8\times
   10^{11}$ & $5.9\times 10^{11}$ &
   $6.\times 10^{11}$ & $6.3\times
   10^{11}$ \\
 $A(x_b)$ & $-4.4\times 10^{-16}$ &
   $-4.4\times 10^{-16}$ & $-4.4\times
   10^{-16}$ & $-4.4\times 10^{-16}$ &
   $-8.9\times 10^{-16}$ \\
 $A(x_c)$ & $4.6\times 10^{-36}$ &
   $3.5\times 10^{-36}$ & $3.3\times
   10^{-36}$ & $3.2\times 10^{-36}$ &
   $2.6\times 10^{-36}$ \\
 $A(x_b)$ & $-7.3\times 10^{-24}$ &
   $-6.\times 10^{-24}$ & $-5.8\times
   10^{-24}$ & $-5.6\times 10^{-24}$ &
   $-5.\times 10^{-24}$ \\
 $A(x_c)$ & $4.7\times 10^{-40}$ &
   $-1.7\times 10^{-39}$ & $-1.\times
   10^{-39}$ & $-7.4\times 10^{-40}$ &
   $8.1\times 10^{-40}$ \\
\hline
\end{tabular}
\center Table 5: Quantum Lensing characteristics for observed most
massive black holes 21-25, $\mu_4=-\mu_2,\mu_3=-\mu_1$
\begin{tabular}{|c|c|c|c|c|c|}
\hline $Black~Hole$ & 21 & 22 &23 & 24 & 25 \\
\hline
 $\epsilon $ & $7.3\times 10^{-24}$ & $5.9\times
   10^{-24}$ & $5.\times 10^{-24}$ &
   $4.8\times 10^{-24}$ & $4.8\times
   10^{-24}$ \\
 $x_b$ & $0.5$ & $0.5$ & $0.5$ & $0.5$ & $0.5$
   \\
 $x_c$ & $7.4\times 10^{11}$ & $8.2\times
   10^{11}$ & $9.\times 10^{11}$ &
   $9.1\times 10^{11}$ & $9.2\times
   10^{11}$ \\
 $A(x_b)$ & $-4.4\times 10^{-16}$ &
   $-4.4\times 10^{-16}$ & $-4.4\times
   10^{-16}$ & $-8.9\times 10^{-16}$ &
   $-8.9\times 10^{-16}$ \\
 $A(x_c)$ & $1.6\times 10^{-36}$ &
   $1.2\times 10^{-36}$ & $9.3\times
   10^{-37}$ & $8.7\times 10^{-37}$ &
   $8.7\times 10^{-37}$ \\
 $A(x_b)$ & $-3.7\times 10^{-24}$ &
   $-2.9\times 10^{-24}$ & $-2.5\times
   10^{-24}$ & $-2.4\times 10^{-24}$ &
   $-2.4\times 10^{-24}$ \\
 $A(x_c)$ & $2.4\times 10^{-40}$ &
   $1.3\times 10^{-40}$ & $2.3\times
   10^{-40}$ & $-2.6\times 10^{-40}$ &
   $1.3\times 10^{-41}$ \\
\hline
\end{tabular}
\center Table 6: Quantum Lensing characteristics for observed most
massive black holes 26-30, $\mu_4=-\mu_2,\mu_3=-\mu_1$
\begin{tabular}{|c|c|c|c|c|c|}
\hline $Black~Hole$ & 26 & 27 & 28 & 29 & 30\\
\hline
 $\epsilon $ & $3.3\times 10^{-24}$ & $3.2\times
   10^{-24}$ & $3.\times 10^{-24}$ &
   $2.7\times 10^{-24}$ & $2.4\times
   10^{-24}$ \\
 $x_b$ & $0.5$ & $0.5$ & $0.5$ & $0.5$ & $0.5$
   \\
 $x_c$ & $1.1\times 10^{12}$ & $1.1\times
   10^{12}$ & $1.2\times 10^{12}$ &
   $1.2\times 10^{12}$ & $1.3\times
   10^{12}$ \\
 $A(x_b)$ & $-4.4\times 10^{-16}$ &
   $-8.9\times 10^{-16}$ & $-4.4\times
   10^{-16}$ & $-8.9\times 10^{-16}$ &
   $-4.4\times 10^{-16}$ \\
 $A(x_c)$ & $5.\times 10^{-37}$ &
   $4.7\times 10^{-37}$ & $4.3\times
   10^{-37}$ & $3.6\times 10^{-37}$ &
   $3.2\times 10^{-37}$ \\
 $A(x_b)$ & $-1.7\times 10^{-24}$ &
   $-1.6\times 10^{-24}$ & $-1.5\times
   10^{-24}$ & $-1.3\times 10^{-24}$ &
   $-1.2\times 10^{-24}$ \\
 $A(x_c)$ & $9.9\times 10^{-41}$ &
   $4.2\times 10^{-41}$ & $-4.8\times
   10^{-40}$ & $2.9\times 10^{-40}$ &
   $1.9\times 10^{-40}$ \\
\hline
\end{tabular}
\center Table 7: Quantum Lensing characteristics for observed most
massive black holes 31-35, $\mu_4=-\mu_2,\mu_3=-\mu_1$
\begin{tabular}{|c|c|c|c|c|c|}
\hline $Black~Hole$ & 31 & 32 & 33 & 34 & 35 \\
\hline
 $\epsilon $ & $2.\times 10^{-24}$ & $1.9\times
   10^{-24}$ & $1.5\times 10^{-24}$ &
   $1.4\times 10^{-24}$ & $1.2\times
   10^{-24}$ \\
 $x_b$ & $0.5$ & $0.5$ & $0.5$ & $0.5$ & $0.5$
   \\
 $x_c$ & $1.4\times 10^{12}$ & $1.4\times
   10^{12}$ & $1.6\times 10^{12}$ &
   $1.7\times 10^{12}$ & $1.9\times
   10^{12}$ \\
 $A(x_b)$ & $-4.4\times 10^{-16}$ &
   $-4.4\times 10^{-16}$ & $-8.9\times
   10^{-16}$ & $-4.4\times 10^{-16}$ &
   $-8.9\times 10^{-16}$ \\
 $A(x_c)$ & $2.4\times 10^{-37}$ &
   $2.2\times 10^{-37}$ & $1.5\times
   10^{-37}$ & $1.4\times 10^{-37}$ &
   $1.\times 10^{-37}$ \\
 $A(x_b)$ & $-1.\times 10^{-24}$ &
   $-9.6\times 10^{-25}$ & $-7.3\times
   10^{-25}$ & $-6.9\times 10^{-25}$ &
   $-5.8\times 10^{-25}$ \\
 $A(x_c)$ & $-1.1\times 10^{-40}$ &
   $1.5\times 10^{-40}$ & $1.6\times
   10^{-41}$ & $1.2\times 10^{-40}$ &
   $-2.\times 10^{-41}$ \\
\hline
\end{tabular}
\center Table 8: Quantum Lensing characteristics for observed most
massive black holes 36-40, $\mu_4=-\mu_2,\mu_3=-\mu_1$
\begin{tabular}{|c|c|c|c|c|c|}
\hline $Black~Hole$ & 36 & 37 & 38 & 39 & 40\\
\hline
 $\epsilon $ & $1.1\times 10^{-24}$ & $8.3\times
   10^{-25}$ & $6.9\times 10^{-25}$ &
   $6.\times 10^{-25}$ & $4.8\times
   10^{-25}$ \\
 $x_b$ & $0.5$ & $0.5$ & $0.5$ & $0.5$ & $0.5$
   \\
 $x_c$ & $1.9\times 10^{12}$ & $2.2\times
   10^{12}$ & $2.4\times 10^{12}$ &
   $2.6\times 10^{12}$ & $2.9\times
   10^{12}$ \\
 $A(x_b)$ & $-4.4\times 10^{-16}$ &
   $-4.4\times 10^{-16}$ & $-4.4\times
   10^{-16}$ & $-4.4\times 10^{-16}$ &
   $-4.4\times 10^{-16}$ \\
 $A(x_c)$ & $9.4\times 10^{-38}$ &
   $6.3\times 10^{-38}$ & $4.8\times
   10^{-38}$ & $3.9\times 10^{-38}$ &
   $2.8\times 10^{-38}$ \\
 $A(x_b)$ & $-5.4\times 10^{-25}$ &
   $-4.2\times 10^{-25}$ & $-3.5\times
   10^{-25}$ & $-3.\times 10^{-25}$ &
   $-2.4\times 10^{-25}$ \\
 $A(x_c)$ & $5.7\times 10^{-41}$ &
   $3.8\times 10^{-41}$ & $-3.3\times
   10^{-41}$ & $1.1\times 10^{-41}$ &
   $1.9\times 10^{-41}$ \\
\hline
\end{tabular}
\center Table 9: Quantum Lensing characteristics for observed most
massive black holes 41-45, $\mu_4=-\mu_2,\mu_3=-\mu_1$
\begin{tabular}{|c|c|c|c|c|c|}
\hline $Black~Hole$ & 41 & 42 & 43 & 44 & 45 \\
\hline
 $\epsilon $ & $2.7\times 10^{-25}$ & $1.2\times
   10^{-25}$ & $9.7\times 10^{-26}$ &
   $1.9\times 10^{-26}$ & $1.5\times
   10^{-26}$ \\
 $x_b$ & $0.5$ & $0.5$ & $0.5$ & $0.5$ & $0.5$
   \\
 $x_c$ & $3.8\times 10^{12}$ & $5.8\times
   10^{12}$ & $6.4\times 10^{12}$ &
   $1.4\times 10^{13}$ & $1.6\times
   10^{13}$ \\
 $A(x_b)$ & $-8.9\times 10^{-16}$ &
   $-4.4\times 10^{-16}$ & $-4.4\times
   10^{-16}$ & $-4.4\times 10^{-16}$ &
   $-8.9\times 10^{-16}$ \\
 $A(x_c)$ & $1.2\times 10^{-38}$ &
   $3.5\times 10^{-39}$ & $2.5\times
   10^{-39}$ & $2.2\times 10^{-40}$ &
   $1.5\times 10^{-40}$ \\
 $A(x_b)$ & $-1.3\times 10^{-25}$ &
   $-6.\times 10^{-26}$ & $-4.9\times
   10^{-26}$ & $-9.6\times 10^{-27}$ &
   $-7.3\times 10^{-27}$ \\
 $A(x_c)$ & $-1.6\times 10^{-41}$ &
   $2.\times 10^{-41}$ & $-5.3\times
   10^{-42}$ & $7.\times 10^{-43}$ &
   $-7.4\times 10^{-43}$ \\
\hline
\end{tabular}
\center Table 10: Quantum Lensing characteristics for observed
most massive black holes 46-50, $\mu_4=-\mu_2,\mu_3=-\mu_1$
\begin{tabular}{|c|c|c|c|c|c|}
\hline $Black~Hole$ & 46 & 47 & 48 & 49 & 50 \\
\hline
 $\epsilon $ & $1.4\times 10^{-26}$ & $8.7\times
   10^{-27}$ & $6.3\times 10^{-27}$ &
   $1.2\times 10^{-27}$ & $7.7\times
   10^{-28}$ \\
 $x_b$ & $0.5$ & $0.5$ & $0.5$ & $0.5$ & $0.5$
   \\
 $x_c$ & $1.7\times 10^{13}$ & $2.1\times
   10^{13}$ & $2.5\times 10^{13}$ &
   $5.8\times 10^{13}$ & $7.2\times
   10^{13}$ \\
 $A(x_b)$ & $-8.9\times 10^{-16}$ &
   $-8.9\times 10^{-16}$ & $-4.4\times
   10^{-16}$ & $-4.4\times 10^{-16}$ &
   $-4.4\times 10^{-16}$ \\
 $A(x_c)$ & $1.4\times 10^{-40}$ &
   $6.8\times 10^{-41}$ & $4.2\times
   10^{-41}$ & $3.5\times 10^{-42}$ &
   $1.8\times 10^{-42}$ \\
 $A(x_b)$ & $-6.9\times 10^{-27}$ &
   $-4.4\times 10^{-27}$ & $-3.2\times
   10^{-27}$ & $-6.\times 10^{-28}$ &
   $-3.8\times 10^{-28}$ \\
 $A(x_c)$ & $-9.4\times 10^{-43}$ &
   $6.4\times 10^{-43}$ & $-1.7\times
   10^{-43}$ & $1.\times 10^{-43}$ &
   $6.\times 10^{-44}$ \\
\hline
\end{tabular}
\center Table 11: Quantum Lensing characteristics for observed
most massive black holes 51-55, $\mu_4=-\mu_2,\mu_3=-\mu_1$
\begin{tabular}{|c|c|c|c|c|c|}
\hline $Black~Hole$ & 51 & 52 & 53 & 54 & 55 \\
\hline
 $\epsilon $ & $5.9\times 10^{-28}$ & $3.6\times
   10^{-28}$ & $2.8\times 10^{-28}$ &
   $1.9\times 10^{-28}$ & $1.1\times
   10^{-28}$ \\
 $x_b$ & $0.5$ & $0.5$ & $0.5$ & $0.5$ & $0.5$
   \\
 $x_c$ & $8.2\times 10^{13}$ & $1.\times
   10^{14}$ & $1.2\times 10^{14}$ &
   $1.4\times 10^{14}$ & $1.9\times
   10^{14}$ \\
 $A(x_b)$ & $-4.4\times 10^{-16}$ &
   $-8.9\times 10^{-16}$ & $-4.4\times
   10^{-16}$ & $-4.4\times 10^{-16}$ &
   $-4.4\times 10^{-16}$ \\
 $A(x_c)$ & $1.2\times 10^{-42}$ &
   $5.8\times 10^{-43}$ & $3.8\times
   10^{-43}$ & $2.2\times 10^{-43}$ &
   $9.4\times 10^{-44}$ \\
 $A(x_b)$ & $-2.9\times 10^{-28}$ &
   $-1.8\times 10^{-28}$ & $-1.4\times
   10^{-28}$ & $-9.6\times 10^{-29}$ &
   $-5.4\times 10^{-29}$ \\
 $A(x_c)$ & $-2.2\times 10^{-44}$ &
   $-4.7\times 10^{-44}$ & $3.9\times
   10^{-44}$ & $3.\times 10^{-44}$ &
   $-1.2\times 10^{-44}$ \\
\hline
\end{tabular}
\center Table 12: Quantum Lensing characteristics for observed
most massive black holes 56-60, $\mu_4=-\mu_2,\mu_3=-\mu_1$
\begin{tabular}{|c|c|c|c|c|c|}
\hline $Black~Hole$ & 56 & 57 & 58 & 59 & 60 \\
\hline
 $\epsilon $ & $6.9\times 10^{-31}$ & $4.8\times
   10^{-29}$ & $1.2\times 10^{-29}$ &
   $3.\times 10^{-30}$ & $2.2\times
   10^{-30}$ \\
 $x_b$ & $0.5$ & $0.5$ & $0.5$ & $0.5$ & $0.5$
   \\
 $x_c$ & $2.4\times 10^{15}$ & $2.9\times
   10^{14}$ & $5.8\times 10^{14}$ &
   $1.2\times 10^{15}$ & $1.3\times
   10^{15}$ \\
 $A(x_b)$ & $-8.9\times 10^{-16}$ &
   $-4.4\times 10^{-16}$ & $-4.4\times
   10^{-16}$ & $-4.4\times 10^{-16}$ &
   $-4.4\times 10^{-16}$ \\
 $A(x_c)$ & $4.8\times 10^{-47}$ &
   $2.8\times 10^{-44}$ & $3.5\times
   10^{-45}$ & $4.3\times 10^{-46}$ &
   $2.8\times 10^{-46}$ \\
 $A(x_b)$ & $-3.5\times 10^{-31}$ &
   $-2.4\times 10^{-29}$ & $-6.\times
   10^{-30}$ & $-1.5\times 10^{-30}$ &
   $-1.1\times 10^{-30}$ \\
 $A(x_c)$ & $7.5\times 10^{-47}$ &
   $6.5\times 10^{-45}$ & $2.2\times
   10^{-45}$ & $4.5\times 10^{-46}$ &
   $-3.\times 10^{-46}$ \\
\hline
\end{tabular}
\center
  Table 13: Quantum deflection angle calculated real
values for $0.001<\epsilon<0.01$ where $\mu_{3}=-\mu_{1},
\mu_{4}=-\mu_{2}$
\begin{tabular}{|c|c|c|c|c|c|c|c|}
  \hline
 $\epsilon$  & $\alpha_{q}$ & $\theta_1$  & $\theta_2$  & $\theta_3$  & $\theta_4$  & $\mu_{1}$ & $\mu_{2}$\\
 \hline
 0.0022 & -0.598212 & -2.36947 & -0.584932 & 0.772118 & 2.55666 &-0.525425 & 1.62159\\
 0.0023 & -0.598235 & -2.3695 & -0.584925 & 0.772088 & 2.55667 & -0.52541 & 1.62165 \\
 0.0025 & -0.598299 & -2.36959 & -0.584905 & 0.772004 & 2.55669 & -0.525365 & 1.62183 \\
 0.0026 & -0.59834 & -2.36964 & -0.584893 & 0.771951 & 2.5567 & -0.525337 & 1.62194\\
 0.0027 & -0.598386 & -2.3697 & -0.584879 & 0.771891 & 2.55671 & -0.525306 & 1.62207\\
 0.0028 & -0.598437 & -2.36977 & -0.584864 & 0.771825 & 2.55673 & -0.525271 & 1.6222 \\
 0.0029 & -0.598492 & -2.36984 & -0.584847 & 0.771753 & 2.55675 & -0.525233 & 1.62235\\
\hline
\end{tabular}
\center Continuation of Table 13
\begin{tabular}{|c|c|c|c|c|c|c|c|}
  \hline
 $\epsilon$  & $\alpha_{q}$ & $\theta_1$  & $\theta_2$  & $\theta_3$  & $\theta_4$  & $\mu_{1}$ & $\mu_{2}$ \\
 \hline
 0.0069 & -0.602593 & -2.37518 & -0.583608 & 0.766414 & 2.55798 & -0.522409 & 1.63366\\
 0.0070 & -0.602717 & -2.37534 & -0.583571 & 0.766252 & 2.55802 & -0.522323 & 1.634 \\
 0.0071 & -0.602841 & -2.3755 & -0.583533 & 0.76609 & 2.55806 & -0.522237 & 1.63435 \\
 0.0073 & -0.60309 & -2.37583 & -0.583458 & 0.765765 & 2.55813 & -0.522065 & 1.63505\\
 0.0074 & -0.603215 & -2.37599 & -0.58342 & 0.765603 & 2.55817 & -0.521979 & 1.6354 \\
 0.0075 & -0.60334 & -2.37615 & -0.583382 & 0.765441 & 2.55821 & -0.521893 & 1.63575 \\
 0.0079 & -0.603838 & -2.3768 & -0.583232 & 0.764792 & 2.55836 & -0.521549 & 1.63714\\
 0.0080 & -0.603961 & -2.37696 & -0.583194 & 0.764631 & 2.5584 & -0.521463 & 1.63749\\
 0.0081 & -0.604085 & -2.37712 & -0.583157 & 0.76447 & 2.55844 & -0.521378 & 1.63784\\
 0.0085 & -0.604574 & -2.37776 & -0.583009 & 0.763834 & 2.55858 & -0.52104 & 1.63922\\
 0.0086 & -0.604694 & -2.37792 & -0.582972 & 0.763676 & 2.55862 & -0.520957 & 1.63956\\
 0.0091 & -0.60528 & -2.37868 & -0.582795 & 0.762914 & 2.5588 & -0.520551 & 1.64122\\
 0.0092 & -0.605393 & -2.37883 & -0.582761 & 0.762766 & 2.55883 & -0.520473 & 1.64154\\
 0.0094 & -0.605613 & -2.37911 & -0.582694 & 0.76248 & 2.5589 & -0.520321 & 1.64216\\
 0.0095 & -0.605719 & -2.37925 & -0.582662 & 0.762341 & 2.55893 & -0.520247 & 1.64246\\
 0.0096 & -0.605823 & -2.37939 & -0.58263 & 0.762205 & 2.55896 & -0.520175 & 1.64276\\
 0.0099 & -0.606115 & -2.37977 & -0.582542 & 0.761825 & 2.55905 & -0.519972 & 1.64359\\
\hline
\end{tabular}
 \center
 Table 14: Classical deflection angle calculated real values
for $0.001<\epsilon<0.01$ where $\mu_{3}=-\mu_{1},
\mu_{4}=-\mu_{2}$
\begin{tabular}{|c|c|c|c|c|c|c|c|c|}
  \hline
 $\epsilon$ & $\alpha _c$ & $\theta _1$ & $\theta _2$ & $\theta _3$ & $\theta _4$ & $\mu_1$ & $\mu_2$ \\
  \hline
 $0.0023$ & $2.443$ & $-2.501$ & $-0.5537$ & $0.6409$ & $2.588$ & $-0.4528$ & $2.009$ \\
 $0.0024$ & $2.441$ & $-2.504$ & $-0.5529$ & $0.6377$ & $2.589$ & $-0.4510$ & $2.022$ \\
 $0.0025$ & $2.439$ & $-2.507$ & $-0.5521$ & $0.6346$ & $2.589$ & $-0.4492$ & $2.036$\\
 $0.0026$ & $2.436$ & $-2.51$ & $-0.5514$ & $0.6316$ & $2.59$ & $-0.4474$ & $2.049$\\
 $0.0027$ & $2.434$ & $-2.513$ & $-0.5506$ & $0.6286$ & $2.591$ & $-0.4457$ & $2.062$\\
 $0.0028$ & $2.432$ & $-2.516$ & $-0.5499$ & $0.6256$ & $2.592$ & $-0.4439$ & $2.076$\\
 $0.0029$ & $2.43$ & $-2.519$ & $-0.5492$ & $0.6228$ & $2.592$ & $-0.4423$ & $2.089$\\
\hline
\end{tabular}
  Continuation of Table 14
\begin{tabular}{|c|c|c|c|c|c|c|c|c|}
  \hline
 $\epsilon$ & $\alpha _c$ & $\theta _1$ & $\theta _2$ & $\theta _3$ & $\theta _4$ & $\mu_1$ & $\mu_2$\\
  \hline
$0.0059$ &$2.376$ &$-2.591$ &$-0.5308$ &$0.5508$ &$2.611$ & $-0.3993$ & $2.534$\\
$0.0060$ &$2.375$ &$-2.593$ &$-0.5303$ &$0.5487$ &$2.611$ & $-0.398$ & $2.55$\\
  $0.0061$ &$2.373$ &$-2.595$ &$-0.5297$ &$0.5467$ &$2.612$ & $-0.3968$ & $2.568$\\
 $0.0062$ &$2.372$ &$-2.597$ &$-0.5292$ &$0.5447$ &$2.612$ & $-0.3955$ & $2.585$\\
$0.0063$ &$2.37$ &$-2.599$ &$-0.5287$ &$0.5426$ &$2.613$ & $-0.3943$ & $2.602$\\
$0.0067$ &$2.364$ &$-2.607$ &$-0.5266$ &$0.5347$ &$2.615$ & $-0.3894$ & $2.675$ \\
$0.0068$ &$2.363$ &$-2.609$ &$-0.526$ &$0.5327$ &$2.616$ & $-0.3882$ & $2.693$\\
$0.0069$ &$2.361$ &$-2.611$ &$-0.5255$ &$0.5308$ &$2.616$ & $-0.387$ & $2.712$\\
$0.0070$ &$2.36$ &$-2.613$ &$-0.525$ &$0.5289$ &$2.617$ & $-0.3858$ & $2.731$\\
$0.0071$ &$2.359$ &$-2.615$ &$-0.5245$ &$0.5269$ &$2.617$ & $-0.3846$ & $2.75$\\
$0.0072$ &$2.357$ &$-2.617$ &$-0.524$ &$0.525$ &$2.618$ & $-0.3834$ & $2.77$\\
$0.0073$ &$2.356$ &$-2.618$ &$-0.5235$ &$0.5231$ &$2.618$ & $-0.3823$ & $2.79$\\
$0.0075$ &$2.353$ &$-2.622$ &$-0.5225$ &$0.5194$ &$2.619$ & $-0.3799$ & $2.83$\\
$0.0076$ &$2.352$ &$-2.624$ &$-0.522$ &$0.5175$ &$2.62$ & $-0.3788$ & $2.85$\\
$0.0078$ &$2.349$ &$-2.628$ &$-0.521$ &$0.5138$ &$2.621$ & $-0.3765$ & $2.892$\\
$0.0079$ &$2.348$ &$-2.63$ &$-0.5205$ &$0.512$ &$2.621$ & $-0.3753$ & $2.914$\\
$0.0080$ &$2.346$ &$-2.6310$ &$-0.5200$ &$0.5102$ &$2.622$ & $-0.3742$ & $2.936$\\
\hline
\end{tabular}
\vspace {0.5cm}\\
{\bf References}
\begin{description}
\item[1] A. O. Petters, H. Levine and
J. Wambsganss, \textit{Singularity theory and Gravitational
lensing}, (Birkhauser, Boston, 2001).
\item[2] P. E. J. Schneider and E. E. Falco \textit{Gravitational lenses}, (Springer, Berlin,
  1992).
  \item[3] P. E. J. Schneider, \textit{Gravitational Lensing: Strong, Weak and Micro} 33,
269, (2006)  Astro-ph/0509252.
\item[4] G. V. Kraniotis, Gen. Rel. Grav. 46, 11,
 1818 (2014).
 \item[5] E. F. Eiroa, G. E. Romero and D. F. Torres, Phys. Rev. D, 66,
  024010 (2002).
 \item[6] M. Sereno, Phys. Rev. D, 77,
043004 (2008).
 \item[7] M. Sereno and F. D. Luca, Phys. Rev. D, 74, 123009
 (2006).
 \item[8] V. Bozza, F. De
Luca and G. Scarpetta, Phys. Rev. D, 74, 063001 (2006).
 \item[9] T. K. Dey and S. Sen, Mod. Phys. Lett. A, 23, 953 (2008); gr-qc/0806.4059
\item[10] G. N.  Gyulchev and S. S. Yazadjiev, Phys. Rev. D, 78, 083004 (2008); gr-qc/0806.3289
\item[11] K. S. Virbhadra,  D. Narasimha and S. M. Chitre, Astronom. Astrophys.
337, 1 (1998).
\item[12] J. Bardeen, Proc. GRS, Tiflis, USSR (1968).
\item[13] E. F. Eiroa and C. M. Sendra, Class. Quantum
Grav. 28, 085008 (2011).
\item[14] S. Ansoldi,  \textit{Apeared
in the proceedings of "BH2, Dynamics and Thermodynamics of Black
holes and Naked Singularities", May 10-12,
 Milano, Italy, 2007},
(gr-qc/0802.0330).
\item[15]  E. Ayon-Beato and A. Garcia, Phys. Lett. B, 493, 149, (2000).
\item[16] H. Ghaffarnejad, H. Neyad and M. A. Mojahedi, Astrophys. Space Sci, 346, 497 (2013); (Physics.gen-ph/1305.6914).
\item[17] S. Weinberg, \textit{Gravitation and Cosmology}, Principles and
Applications of the General Theory of Relativity (John Wiley and
Sons, Inc. New York 1972).
\item[18] E. F. Eiroa, Phys. Rev. D, 71, 083010 (2005).
\item[19] V. Bozza,  Phys. Rev. D, 78, 103005, (2008).
 \item[20] C. R. Keeton and A. O. Petters, Phys. Rev. D, 72,
 104006 (2005).
\item[21] M. Casals, A. Fabbri, C. Martinez and J. Zanelli, gr-qc/1608.05366v1 (2016).
\item[22] M. Banados, C. Teitelboim and J. Zanelli Phys. Rev. Lett. 69,
 1849 (1992).
\item[23] S. Carlip,  gr-qc/9506079 (1995).
\item[24] B. Gaensler \textit{Extreme Cosmos: A
Guided Tour of the Fastest, Brightest, Hottest, Heaviest, Oldest,
and Most Amazing Aspects of Our Universe. 2012} ISBN
978-1-101-58701-0.
\item[25] G. Ghisellini, L. Foschini, M. Volonteri, G. Ghirlanda, F. Haardt, D. Burlon, F. Tavecchio et al.
, Month. Notices Roy. Astronom. Soc. Lett. v2. 399: L24 (2009).
arXiv:0906.0575.
\item[26] G. Ghisellini, R. D. Ceca, M. Volonteri, G. Ghirlanda, F. Tavecchio, L. Foschini, G. Tagliaferri, F. Haardt, G. Pareschi, J. Grindlay,
 Month. Notices Roy. Astronom. Soc. 405: 387. (2010),
 arXiv:0912.0001.
\item[27] W. Zuo, X. B. Wu, X. Fan, R.  Green, R. Wang, F. Bian, Astrophys. J. 799 (2): 189 (2014), arXiv:1412.2438.
\item[28] S. A. Walker, A. C. Fabian, H. R. Russell, J. S. Sanders, Month. Notices Roy. Astronom. Soc. 442 (3): 2809 (2014), arXiv:1405.7522v1
 [astro-ph.HE].
 \item[29] F. L. Geraint, C. S. Chapman, A. I. Rodrigo, J. I. Michael and J. T. Edward,  Astrophys. J. Lett., 505,
 1 (1998).
 \item[30]  J. N. McConnell, M. C. Pei, K. Gebhardt, A. W. Shelley, D. M. Jeremy,  R. L. Tod, R. J. Graham, O. D. Richstone,
   Nature. 480, 7376, 215-8 (2011). arXiv:1112.1078
\item[31] M. McDonald, M. Bayliss, B. A. Benson, R. J. Foley, J. Ruel, P. Sullivan, S. Veilleux, K. A. Aird,
  M. L. N. Ashby, M. Bautz, G. Bazin,   L. E. Bleem, M. Brodwin, J. E. Carlstrom, C. L. Chang, H. M. Cho, A. Clocchiatti, T. M. Crawford, A. T. Crites,
  T.  D. Haan, S. Desai, M. A. Dobbs, J. P. Dudley, E. Egami, W. R. Forman, G. P. Garmire, E. M. George, M. D. Gladders, A. H. Gonzalez
   et al., Nature. 488, 7411:
349-52 (2012), arXiv:1208.2962
\item[32] M. J. Valtonen, S. Ciprini, H. J. Lehto, Month. Notices Roy. Astronom. Soc. 427: 77 (2012).
  arXiv:1208.0906
\item[33]\textit{http://www.nasa.gov/feature/goddard/2016/behemoth-black-hole-found-in-an-unlikely-place}
 \item[34] X. Wu, F. Wang, X. Fan, W. Yi, W. Zuo,; F. Bian, L. Jiang,
I. D. McGreer, R. Wang, J. Yang, Q. Yang, D. Thompson, Y.
Beletsky,
 Nature. 518, 7540, 512-515 (2015), arXiv:1502.07418 .
 \item[35] G. Ghisellini G. Tagliaferri, T. Sbarrato, N. Gehrels,  Month. Notices Roy. Astronom. Soc. Lett. 450, L34 (2015)
  . arXiv:1501.07269
\item[36] \textit{ Most Powerful Eruption In The Universe
Discovered NASA/Marshall Space Flight Center (ScienceDaily)
January 6, 2005}
\item[37]  L. Tingting, G. Suvi, H. Sebastien, A. M. Eugene, S. W.
Burgett, K.  Chambers, H. Flewelling, M. Huber, W. K. Hodapp, N.
Kaiser, R. P. Kudritzki, J. L. Tonry, R. J. Wainscoat, C. Waters,
Astrophys. J.  803 (2): L16 (2015). arXiv:1503.02083
\item[38] J. H. Larrondo,  S. W. Allen, G. B. Taylor, A. C.
 Fabian, R. E. A. Canning, N. Werner, J. S. Sanders, C. K. Grimes, S. Ehlert, A. V. D. Linden, Astrophys. J. 777 (2):
 163 (2013). arXiv:1306.0907
\item[39] O. Lopez-Cruz; C. Anorve, M. Birkinshaw, D. M. Worrall, H. J.
  Ibarra-Medel,  W. A. Barkhouse, J. P. Torres-Papaqui, V. Motta,   Astrophys. J.
    795 (2): L31 (2014). arXiv:1405.7758
\item[40] B. Trakhtenbrot, U. C. Megan, F. Civano, D. J. Rosario, M. Elvis, K. Schawinski,
S. Hyewon, A. Bongiorno, B. D Simmons, Science, 349, 168 (2015).
arXiv:1507.02290
\item[41] J. L. Walsh, A. J. Barth, L. C. Ho, M. Sarzi, Astrophys. J.   770 (2): 86 (2013), arXiv:1304.7273
\item[42] E. Emsellem, Month.Notices. Roy. Astronom. Soc.  433 (3), 1862, (2013).
 arXiv:1305.3630.
\item[43] J. Shen and K. Gebhardt, Astrophys. J. 711: 484-494 (2010), arXiv:0910.4168 .
\item[44] C. Alberto, S. Morrison and P. Morrison,
Astronom. J., 123,5 (2002).
\item[45] J. Matson \textit{"Brilliant, but
Distant: Most Far-Flung Known Quasar Offers Glimpse into Early
Universe". Scientific American. 2011}
\item[46] D. J. Mortlock,  S. J. Warren, B. P. Venemans,  Patel, Hewett, McMahon, Simpson, Theuns, Gonzales-Solares, Adamson, Dye, Hambly, Hirst,
Irwin, Kuiper, Lawrence, Rottgering, et al.,
 Nature. 474
,7353,: 616-619 (2011). arXiv:1106.6088.
\item[47] J. Kormendy, D. Richstone,   Astrophys. J. 393: 559-578 (1992).
 \item[48] R. W. Romani, Astronom. J.
    132 (5): 1959-1963 (2006). arXiv:astro-ph/0607581
\item[49]  G. A. Bower, Astrophys. J. 492 (1):
111-114, (1998); arXiv:astro-ph/9710264.
\item[50] R. Bender, J. Kormendy, M. E. Cornel and D. B. Fisher,
Astro-Phys. GA/1411.2598v1 (2014).
 \item[51]  \textit{"Black
Holes: Gravity's Relentless Pull interactive: Encyclopedia".},
HubbleSite, (2015).
\item[52] J. Kormendy, R. Bender, E. A. Ajhar,  A.
Dressler, S. M. Faber, K. Gebhardt, C. Grillmair, T. R. Lauer, D.
Richstone, S. Tremaine,  Astrophys. J. Lett. 473 (2): L91-L94
(1996).
\item[53] M. Loewenstein,
  Astrophys. J. 555 (1): L21-L24 (2001). arXiv:astro-ph/0106326 .
\item[54]  \textit{"Astronomers
Discover Record-Breaking Quasar"}, Sci-News.com. (2015).
\end{description}
\end{document}